# Gamma-ray Emission from Globular Clusters

Pak-Hin T. Tam[1,2†], Chung Y. Hui[3], Albert K. H. Kong[2]

[1]Institute of Astronomy and Space Science, Sun Yat-Sen University, Guangzhou 510275, China
[2]Institute of Astronomy and Department of Physics, National Tsing Hua University, Hsinchu 30013, Taiwan
[3]Department of Astronomy and Space Science, Chungnam National University, Daejeon 34134, Korea

Over the last few years, the data obtained using the Large Area Telescope(LAT) aboard the Fermi Gamma-ray Space Telescope has provided new insights on high-energy processes in globular clusters, particularly those involving compact objects such as MilliSecond Pulsars (MSPs). Gamma-ray emission in the 100 MeV to 10 GeV range has been detected from more than a dozen globular clusters in our galaxy, including 47 Tucanae and Terzan 5. Based on a sample of known gamma-ray globular clusters, the empirical relations between gamma-ray luminosity and properties of globular clusters such as their stellar encounter rate, metallicity, and possible optical and infrared photon energy densities, have been derived. The measured gamma-ray spectra are generally described by a power law with a cut-off at a few gigaelectronvolts. Together with the detection of pulsed $\gamma$-rays from two MSPs in two different globular clusters, such spectral signature lends support to the hypothesis that $\gamma$-rays from globular clusters represent collective curvature emission from magnetospheres of MSPs in the clusters. Alternative models, involving Inverse-Compton (IC) emission of relativistic electrons that are accelerated close to MSPs or pulsar wind nebula shocks, have also been suggested. Observations at >100 GeV by using Fermi/LAT and atmospheric Cherenkov telescopes such as H.E.S.S.-II, MAGIC-II, VERITAS, and CTA will help to settle some questions unanswered by current data.

## 1. INTRODUCTION

Globular clusters are the oldest gravitationally bounded stellar systems in the Galaxy. Nearly 160 globular clusters are known today (Harris 1996, 2010 edition). These clusters form a spherical halo around the Galaxy, and many of them are located within the Galactic bulge. Owing to the high concentration of stars within globular clusters, these clusters host a large number of compact objects, including neutron stars and White Dwarfs (WDs) many such objects are found in binary systems, forming for example Low-Mass X-ray Binaries (LMXBs) and cataclysmic variables.

Since the 1970s, it has been known that the formation rate per unit mass of LMXBs (Alpar et al. 1982) is several orders of magnitude higher in globular clusters than in the rest of the Galaxy (Clark 1975; Katz 1975). Theoretical arguments have long asserted that the formation of LMXBs is made efficient through frequent stellar encounters. Using the X-ray populations in various globular clusters unveiled by the Chandra X-Ray Observatory, Pooley et al. (2003) and Gendre et al. (2003) found a positive correlation between the number of LMXBs in globular clusters and the stellar encounter rate, $\Gamma_c$, putting the dynamical formation scenario of LMXBs in globular clusters on an observational ground. Because MilliSecond Pulsars (MSPs) are generally believed to be descendants of LMXBs, it is quite natural that ~80% of the known MSPs are detected in globular clusters (cf. Manchester et al. 2005). Recent observations of a class of transitional MSPs further support a strong evolutionary link between LMXBs and radio MSPs (see, e.g., Stappers et al. 2014; Takata et al. 2014; Tam et al. 2014).

Using the cumulative luminosity distribution functions of radio MSPs in globular clusters as a probe of the MSP population in the clusters, Hui et al. (2010) found that the number of MSPs in a globular cluster is correlated with its stellar encounter rate, as well as its metallicity. This finding provides an observational evidence of the dynamical origin of MSPs in globular clusters.

Similar to young pulsars, MSPs can be efficient $\gamma$-ray emitters (Abdo et al. 2009c). Given the large numbers of MSPs that reside in globular clusters, it has been expected that globular clusters are $\gamma$-ray emitters as well, as was first suggested by Chen (1991). In this paper, we review the main results of high-energy $\gamma$-ray observations (100 MeV to ~100 GeV) of Galactic globular clusters, which were mainly contributed by data acquired by using the Large Area Telescope (LAT) on board the Fermi Gamma-ray Space Telescope. Observations at even higher energies (i.e., >100 GeV) by using Cherenkov telescopes such as H.E.S.S. also complement the above mentioned GeV observations and will be presented at the end of this article. The interested reader is referred to an earlier review that presents a more theoretical point of view on $\gamma$-ray emission from globular clusters (Bednarek 2011).

## 2. GEV $\gamma$-RAY OBSERVATIONS FROM GLOBULAR CLUSTERS

### 2.1 Pre-Fermi Era

Observations by using the Energetic Gamma-Ray Experiment Telescope (EGRET) aboard the Compton Gamma-ray Observatory in the 1990s failed to detect any globular clusters, including 47 Tucanae that hosts the largest number of MSPs known at that time (Michelson et al. 1994; Manandhar et al. 1996). Recent LAT observations revealed that the upper limit on the EGRET flux for 47 Tucanae is only twofold higher than the measured LAT flux.

### 2.2 First Discoveries: 47 Tucanae and Terzan 5

The launch of the Fermi Gamma-ray Space Telescope in 2008 has enabled the discovery of $\gamma$-ray emission from globular clusters. The first globular cluster for which $\gamma$-ray emission was discovered was 47 Tucanae (Abdo et al. 2009b, see Fig. 1). When this discovery was announced, MSPs as a class were being established as $\gamma$-ray emitters (Abdo et al. 2009c), and MSPs were believed to be the only stable known γ-ray sources in globular clusters. The observed γ-rays from the direction of 47 Tucanae were thus naturally attributed to the MSPs in the cluster (Abdo et al. 2009b). This assertion was strengthened by the measured cutoff at ~2.5 GeV in the spectrum (Fig. 1) that is close to the average cutoff energy obtained for nearby, γ-ray emitting MSPs in the Galactic field (Abdo et al. 2009c). From the mean γ-ray luminosity of individual MSPs in the Galactic field, i.e., ~5×10$^{33}$ $erg$ $s^{-1}$ (Abdo et al. 2009c), an average MSP would not be detected at distances of several kiloparsecs. It is therefore generally believed that ā-rays from globular clusters do not come from a single MSP, but represent collective emission that originates from the cluster population of MSPs as a whole. That being said, the very luminous PSR J1823–3021A and PSR B1821-24 are exceptional (Freire et al. 2011; Wu et al. 2013). Please also refer to Section 2.4 for more details. Assuming that the average γ-ray luminosity of MSPs in 47 Tucanae is similar to that of nearby MSPs, the number of MSPs that yield the observed flux from 47 Tucanae (~2.5×10$^{-11}$) $erg$ $cm^{-2}$ $s^{-1}$; Abdo et al. 2010b) is ~50, nearly twofold higher than 23 MSPs in 47 Tucanae detected in radio surveys . Such discrepancy may stem from incomplete radio surveys of MSPs, e.g., owing to the viewing angle effect.
The second globular cluster that was found to emit γ-rays was Terzan 5 (Kong et al. 2010, see Fig. 1). Terzan 5 hosts the largest number of detected MSPs up to this date (i.e., 33). Its high-energy spectrum also exhibits a cutoff at ~3 GeV (Abdo et al. 2010b; Kong et al. 2010; see Fig. 1). The observed energy flux above 100 MeV is ~7.1×10$^{-11}$ $erg$ $cm^{-2}$ $s^{-1}$ (Abdo et al. 2010b).

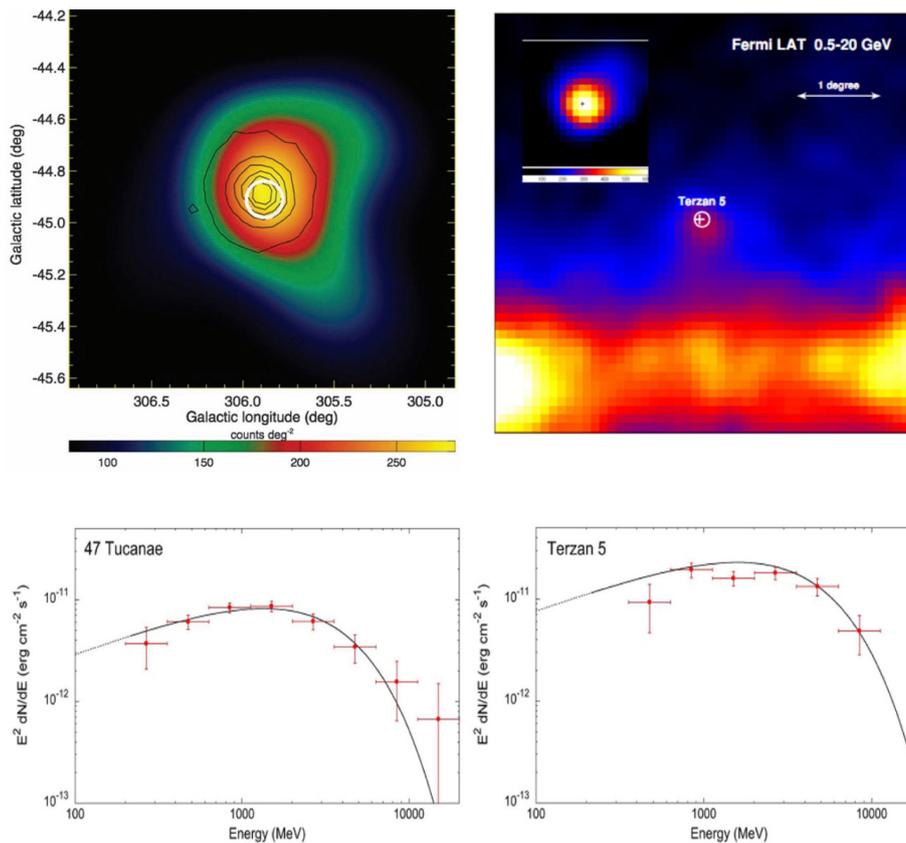

**Fig. 1.** Photon count maps of 47 Tucanae (upper left; Abdo et al. 2009b) and Terzan 5 (upper right; Kong et al. 2010). Their $\gamma$–ray spectra are also shown (Abdo et al. 2010b).

Taking the distance estimates of 47 Tucanae and Terzan 5 to be 4.0 kpc (McLaughlin et al. 2006) and 5.5 kpc (Ortolani et al. 2007), respectively, the γ-ray luminosity of Terzan 5 is roughly fivefold that of 47 Tucanae, indicating that either the actual number of MSPs is higher in Terzan 5 compared with 47 Tucanae, or the average γ-rays efficiency of the underlying MSPs in Terzan 5 is higher than that in 47 Tucanae, or a combination of both. There are some indications from cumulative radio luminosity distributions that Terzan 5 may indeed contain more MSPs than 47 Tucanae (Hui et al. 2010). If the number of MSPs that reside in Terzan 5 is smaller than ~5 times that of 47 Tucanae, environmental factors such as the background soft photons (whose number is much higher in Terzan 5 than in 47 Tucanae), as suggested by Bednarek & Sitarek (2007) and Cheng et al. (2010), may explain the higher γ-ray luminosity of Terzan 5.

### 2.3 Globular Clusters as a Population of γ-ray Sources

After these two initial discoveries, the number of γ-ray emitting globular clusters has continued to grow (e.g., Abdo et al. 2010b; Tam et al. 2011; Zhou et al. 2015). These new findings have established globular clusters as a population of γ-ray sources, and in turn allow one to perform population studies (see Section 3). The Fermi LAT 4-year Point Source Catalog (3FGL) includes 15 sources that are categorized as globular clusters, which do not include two globular clusters for which pulsed γ-rays were observed (see Section 2.4). These 15 sources are listed in Table 1.

**Table 1.** The 15 γ–ray emitting globular clusters included in the 3FGL catalog

| Cluster Name | γ-ray position (J2000) R. A. (h m s) | Dec. (° ´ ˝) | Core Radius (´) | Half-mass Radius (´) | Significance of Detection (σ) | Spectral Curvature (σ) | Energy Flux ($10^{12}$ $erg$ $cm^{-2}$ $s^{-1}$) | Discovery Year | Ref. | # of MSPs |
|---|---|---|---|---|---|---|---|---|---|---|
| 47 Tucanae | 00 23 55.7 | -72 03 55 | 0.36 | 3.17 | 46.4 | 10.4 | 26.1±1.3 | 2009 | (1) | 23 |
| Terzan 5 | 17 48 02.9 | -24 47 15 | 0.15 | 0.52 | 31.6 | 9.4 | 73.1±4.4 | 2010 | (2) | 33 |
| NGC 6388 | 17 36 13.2 | -44 47 15 | 0.12 | 0.52 | 17.0 | 3.3 | 26.5±1.8 | 2010 | (3) | 0 |
| NGC 6440 | 17 48 56.5 | -20 21 16 | 0.14 | 0.48 | 15.1 | 1.6 | 41.3±3.0 | 2010 | (3) | 6 |
| NGC 6541 | 18 07 33.1 | -43 43 09 | 0.18 | 1.06 | 4.9 | 2.9 | 5.6±1.2 | 2010 | (3) | 0 |
| NGC 6652 | 18 35 46.7 | -32 58 14 | 0.10 | 0.48 | 8.1 | 4.6 | 5.8±1.0 | 2010 | (3) | 1 |
| ω Centauri | 13 26 46.8 | -47 27 53 | 2.37 | 5.00 | 11.3 | 6.2 | 9.6±1.3 | 2010 | (4) | 0 |
| M 62 | 17 01 17.5 | -30 06 14 | 0.22 | 0.92 | 18.9 | 6.1 | 20.8±1.9 | 2010 | (4) | 6 |
| NGC 6752 | 19 10 46.8 | -60 00 02 | 0.17 | 1.91 | 8.0 | 3.5 | 5.9±0.9 | 2011 | (5) | 5 |
| M 80 | 16 16 52.6 | -23 00 39 | 0.15 | 0.61 | 5.1 | 2.4 | 6.7±1.5 | 2012 | (6) | 0 |
| 2MS-GC01 | 18 08 33.9 | -19 52 19 | 0.85 | 1.65 | 8.8 | 5.0 | 44.4±6.9 | 2012 | (6) | 0 |
| NGC 2808 | 09 12 17.1 | -64 52 57 | 0.25 | 0.80 | 5.2 | 1.5 | 4.8±1.0 | 2015 | (7) | 0 |
| NGC 6361 | 17 16 38.7 | -28 12 18 | 0.17 | 0.65 | 7.9 | 2.9 | 17.6±2.3 | 2015 | (7) | 0 |
| NGC 6441 | 17 50 15.4 | -37 04 02 | 0.13 | 0.57 | 10.4 | 2.7 | 18.7±2.0 | 2015 | (7) | 4 |
| NGC 6717 | 18 55 11.0 | -22 43 01 | 0.08 | 0.68 | 4.3 | 2.9 | 3.8±1.1 | 2015 | (7) | 0 |

**Notes.** Values in columns 2, 3, 6, 7, & 8, are taken from the 3FGL catalog. Those from columns 4 & 5 are taken from Harris (1996; 2010 edition). Column 9 shows the year of publication that firstly reports a >5 detection of a particular globular cluster. Column 10 lists the corresponding publication: (1) Abdo et al. (2009a); (2) Kong et al. (2010); (3) Abdo et al. (2010a); (4) Abdo et al. (2010b); (5) Tam et al. (2011); (6) Nolan et al. (2012); (7) Acero et al. (2015). Column 11 shows the number of known MSPs as given in http://www.naic.edu/‹···pfreire/GCpsr/.

3
http://janss.kr

**2.4 Searching for γ-ray Pulses from Globular Clusters**

Attempts have been made to detect pulsed γ-rays from individual MSPs in globular clusters. In the first two years after the launch of the Fermi Gamma-ray Space Telescope, no γ-ray pulsation was found by using available ephemerides for MSPs in various globular clusters (Abdo et al. 2009b, 2010b).

The first successful effort to detect γ-ray pulsation from an individual MSP in a globular cluster was made by Freire et al. (2011). They found a significant, 7-σ detection of γ-ray pulsations above 100 MeV from PSR J1823-3021A in the globular cluster NGC 6624 (Fig. 2). The high measured flux in the pulsed component significantly constrained the total number of MSPs in NGC 6624 to be smaller than 32.

Another γ-ray pulsation was detected from the MSP PSR B1821-24 that resides in the globular clusters M28 (Wu et al. 2013; Fig. 3). It has a spin-down power of $\dot{E}=2.2\times10^{36}$ $erg$ $s^{-1}$, which makes it the most energetic MSP detected so far. The significance of pulsed γ-ray emission was reported to be ~4.3σ at energies > 200 MeV (Wu et al. 2013). The pulse profile is shown in Fig. 3. The γ-ray pulsation was finally established by Johnson et al. (2013). A source slightly offset from the M28 core (detected when using off-pulse data of PSR B1821-24) was reported, but its association with M28 remains uncertain (Johnson et al. 2013).

## 3. WHICH PROPERTIES OF GLOBULAR CLUSTERS DETERMINE THE γ-RAY LUMINOSITY?

There have been attempts to relate the γ-ray luminosity of a globular cluster to other properties of the cluster. Such correlations shed light on the origin of γ-rays emitted from globular clusters.

**3.1 Properties of Globular Clusters Related to the γ-ray Emission**

The two-body stellar encounter rate has long been believed to relate to the binary formation rate and hence the number of MSPs in a globular cluster (see, e.g., Pooley et al. 2003; Hui et al. 2010). The first attempt to correlate γ-ray luminosity with the stellar encounter rate was performed by Abdo et al. (2010b). Using the first 8 known γ-ray globular clusters and the expression $\Gamma_c=\rho_0^{\frac{3}{2}}r_c^2$, where $\rho_0$ is the central luminosity density and $r_c$ is the core radius (Verbunt & Hut 1987), they found that the γ-ray luminosity of a globular cluster is correlated with the stellar encounter rate, with a linear correlation coefficient of 0.7. Assuming that each MSP emits a similar amount of γ-rays, the number of MSPs in these γ-ray emitting globular clusters is said to be consistent with the number obtained by other estimators (Abdo et al. 2010b).

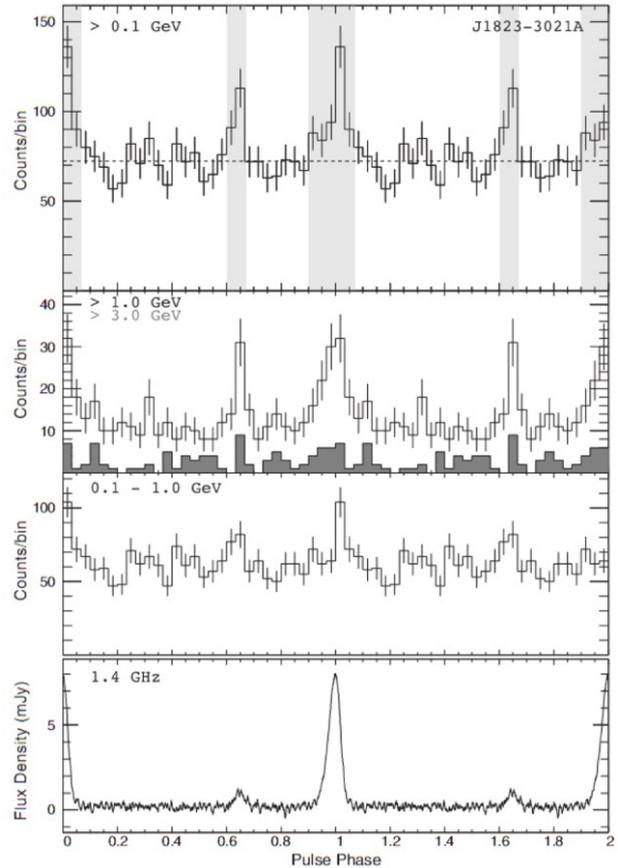

**Fig. 2.** Gamma-ray and radio pulse profiles of PSR J1823-3021A in the globular cluster NGC 6624 (Freire et al. 2011).

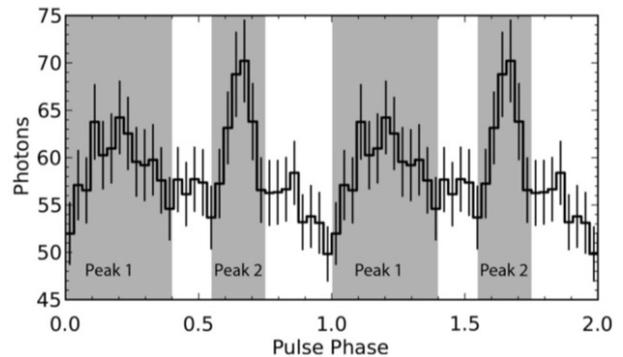

**Fig. 3.** Gamma-ray pulse profile of PSR B1821-24 in the globular cluster M28 (Wu et al. 2013).

Additional parameters besides the stellar encounter rate are thought to relate to the number of MSPs in a globular cluster. Hui et al. (2010) identified the metallicity of a globular cluster as another important indicator of the number of MSPs. A binary system in a globular cluster with higher metallicity is more efficient in orbital shrinkage by magnetic braking. This yields a higher likelihood of a successful Roche-lobe overflow (Ivanova 2006) and in turn leads to a higher formation rate of MSPs. Hui et al. (2010) found a positive correlation between metallicity and the population of MSPs. Using the available γ-ray data from 15 globular clusters, Hui et al. (2011) found that the γ-ray luminosity of a globular cluster is correlated with the cluster metallicity, [Fe/H], at 99.9% level, as log$L_\gamma$ = (0.59±0.15)[Fe/H]+(35.56±0.15).

**Table 2.** Correlation and 1-D regression analysis of $\log L_\gamma$ versus the updated sample of γ–ray globular clusters and their properties

| Parameters | Spearman rank | Prob$_S^a$ | Pearson's $r$ | Prob$_P^b$ | $m^c$ | $c^c$ |
|---|---|---|---|---|---|---|
| $\log \Gamma_c$ | 0.657143 | 0.989338 | 0.612912 | 0.980223 | $0.59 \pm 0.06$ | $34.26 \pm 0.10$ |
| [Fe/H] | 0.802198 | 0.999444 | 0.813197 | 0.999595 | $0.87 \pm 0.06$ | $35.74 \pm 0.05$ |
| $\log u_{\text{optical}}$ | 0.569231 | 0.966381 | 0.532593 | 0.950092 | $0.72 \pm 0.05$ | $34.81 \pm 0.03$ |
| $\log u_{\text{IR}}$ | 0.696703 | 0.994371 | 0.508967 | 0.936935 | $1.11 \pm 0.08$ | $35.47 \pm 0.03$ |

[a] The probability that the Spearman rank correlation coefficient is different from zero.
[b] The probability that the linear correlation coefficient (i.e., Pearson's r) is different from zero.
[c] The best-fit for log $L_\gamma$ = $mx + c$, where $x$ is the corresponding parameter listed in column 1.

On the other hand, primordially formed binaries are not related to the stellar encounter rate. If they form the majority of binaries, one would expect the binary population to be correlated with the cluster mass parameter, $M_{GC}$. Assuming a constant mass-to-light ratio, $M_{GC}$ can be estimated from the absolute visual magnitude $M_V$ according to $M_{GC}=10^{-4M_V}$) (e.g., Hui et al. 2010). This estimate of the cluster mass is advantageous because it is not correlated with the encounter rate, thus serving as an independent parameter in correlational studies. However, no correlation of $M_V$ with the γ-ray luminosity was found (Hui et al. 2011).

Apart from the above three cluster parameters, the energy densities of optical ($u_{opt}$) and infrared ($u_{IR}$) photons at the cluster location are also important for predicting the γ-ray properties, because they serve as the seed photons in IC models (Cheng et al. 2010). Using the GALPROP code (Strong & Moskalenko 1998) to estimate the soft photon energy densities, the γ-ray luminosity was found to correlate with the optical and infrared photon densities, respectively, at >96% confidence level, as $L_\gamma \propto u_{opt}^{0.78\pm0.27}$ as well as $L_\gamma \propto u_{IR}^{1.29\pm0.44}$ (Hui et al. 2011).

With the updated sample of γ-ray globular clusters compiled in the 3FGL catalog (Acero et al. 2015; cf. Table 1), we have re-examined the above correlations by using the revised catalog of globular clusters (Harris 1996; 2010 edition). The corresponding soft photon densities were also estimated by using GALPROP. NGC 6624 and M28 were excluded from all these revised statistical analyses because their γ-ray emissions are dominated by the individual energetic pulsars (see Section 2.4). The results are summarized in Table 2. We found that the correlations remained robust. As indicated by non-parametric analysis (i.e., Spearman rank correlation coefficient), which is more robust than the analysis of linear correlations, the correlations of log$L_\gamma$-[Fe/H] and log$L_\gamma$-log $u_{IR}$ are the strongest among the tested parameters. The best fits for all log $L_\gamma$ data are displayed in Fig. 4.

### 3.2 Fundamental Planes of γ-ray Emission from Globular Clusters

As mentioned above, the total energy output of γ-rays indeed scales with two factors that are crucial to the formation of MSPs: the stellar encounter rate and metallicity. This suggests an intimate relationship between the observed γ-ray emission from globular clusters and the LMXB population, which are precedents of the population of MSPs. The new findings that optical and infrared photon energy densities are also correlated with $L_\gamma$ conform with the IC model (Cheng et al. 2010), in which both the number of MSPs in a globular cluster and the soft photon energy densities play important roles in the resulting γ-ray output. This prompted Hui et al. (2011) to combine two of the above cluster parameters and to perform a 2-dimensional regression analysis.

In view of the updated sample of γ-ray globular clusters and their physical properties (see Section 3.1), we have re-visited these fundamental planes relationships, and the results are given as follows:

$$\log L_\gamma = (34.25 \pm 0.10) + (0.40 \pm 0.07) \log \Gamma_c \\ + (0.63 \pm 0.06) \log u_{opt} \qquad (1)$$

$$\log L_\gamma = (34.80 \pm 0.11) + (0.41 \pm 0.06) \log \Gamma_c \\ + (0.98 \pm 0.09) \log u_{IR} \qquad (2)$$

$$\log L_\gamma = (35.33 \pm 0.07) + (0.61 \pm 0.07)[Fe/H] \\ + (0.50 \pm 0.06) \log u_{opt} \qquad (3)$$

$$\log L_\gamma = (35.79 \pm 0.05) + (0.62 \pm 0.07)[Fe/H] \\ + (0.76 \pm 0.09) \log u_{IR} \qquad (4)$$

The edge-on views of the best-fit fundamental-plane relationships are shown in Fig. 5. These relationships should be taken into account in any realistic model of $\gamma$-ray emission from globular clusters.

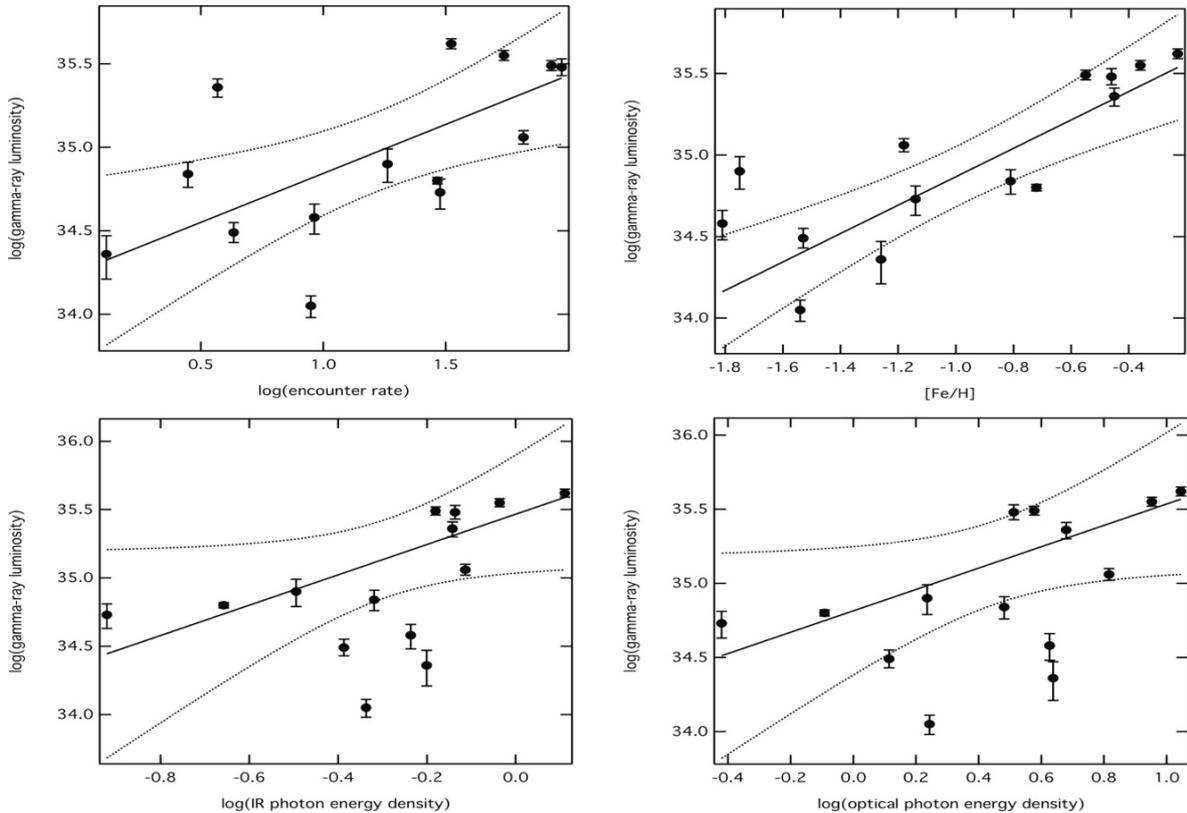

**Fig. 4.** The updated relations between $L_\gamma$ and various individual cluster properties (Acero et al. 2015; Harris 1996). The straight lines in the plots represent the best-fits from the linear regression with the errors of the data points fully taken into account. The dotted lines represent the upper and the lower 95% confidence bands.

## 4. MODELS FOR $\gamma$-RAY EMISSION FROM GLOBULAR CLUSTERS

Two classes of models of $\gamma$-ray emission from globular clusters, both attributed to MSPs, have been discussed in the literature. In pulsar magnetosphere models, $\gamma$-ray emission is produced via curvature radiation within the magnetospheres of MSPs in the clusters (e.g., Venter & de Jager 2008; Venter et al. 2009). The pulsed emission of a number of MSPs is superposed at different frequencies such that unless a small number of luminous MSPs dominate the combined $\gamma$-ray emission, as in the cases for NGC 6624 and M28, pulsations from individual pulsars are difficult to detect.

In another class of models that involve IC emission, electrons are accelerated close to the MSPs or (re-) accelerated in colliding wind shocks and scatter off the optical, infrared, or cosmic microwave background photons (Bednarek & Sitarek 2007; Cheng et al. 2010), giving rise to the observed GeV emission. In this scenario, $\gamma$-rays are intrinsically unpulsed.

The $\gamma$-ray spectrum and especially the cutoff energy as observed for a number of globular clusters are generally similar to nearby MSPs in the Galactic field. Such coincidence may suggest that the observed $\gamma$-rays represent curvature radiation from pulsar magnetospheres. On the other hand, averaged spectra may not be good discriminators for testing various models, because IC models can explain the GeV spectra equally well (Cheng et al. 2010). Moreover, according to Cheng & Taam (2003), MSPs in globular clusters may be very different from those in the Galactic field based on their radio and X-ray properties. In fact, the former may possess complicated magnetic fields owing to frequent stellar encounters (Cheng & Taam 2003), thereby strongly affecting the polar/outer gap structures and in particular quenching the outer gap. This scenario is supported by the fact that the majority of the MSPs in 47 Tucanae are thermal X-ray emitters (Bogdanov et al. 2006).

The differences between the emission signatures of the two models should provide a diagnostic. In the pulsar magnetosphere models, $\gamma$-rays with energies of up to a few gigaelectronvolts are emitted, while IC processes may give rise to $\gamma$-rays with higher energies, reaching teraelectronvolts, depending on the energy of seed photons. A candidate globular cluster reported to emit >10 GeV emission is Liller 1 (Tam et al. 2011). However, this globular cluster does not exist in, e.g., the 3FGL catalog. In addition, no globular clusters have been detected in the 1FHL (Ackermann et al. 2013) and 2FHL catalogs, which include $\gamma$-ray sources detected by the LAT above 10 GeV and 50 GeV, respectively, thus constraining any strong emission above ~10 GeV as predicted by some IC models. Very High Energy (VHE) observations by using Cherenkov telescopes provide further constraints on models at >100 GeV energies (see Section 5).

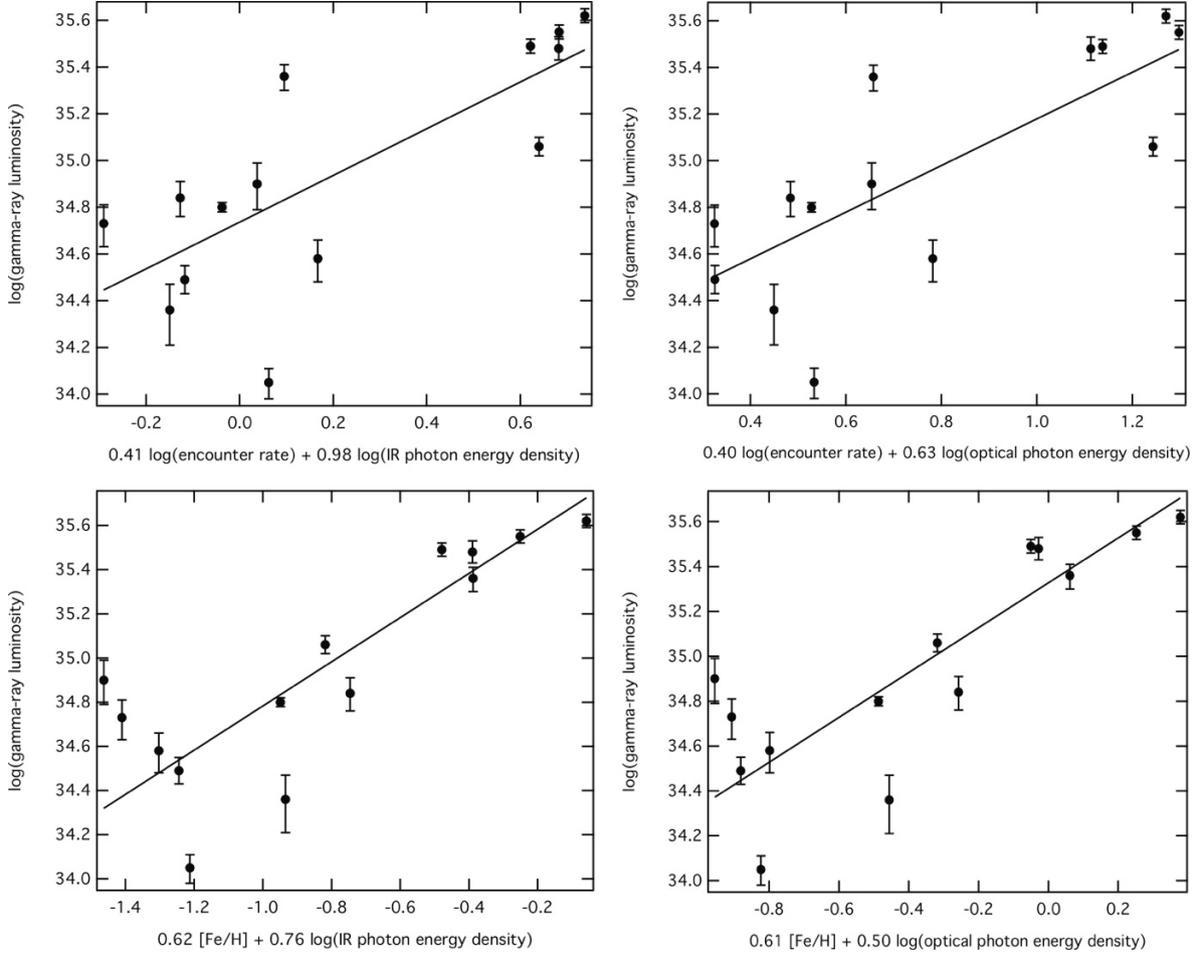

**Fig. 5.** The edge-on views of the fundamental plane relations of ray globular clusters based on the updated sample and the cluster properties (Acero et al. 2015; Harris 1996).

The size of the emission region also differs between the two classes of models. The $\gamma$-ray emission region predicted by the pulsar magnetosphere models is more compact and that predicted by the IC models is more extended. The observed $\gamma$-ray emission region is consistent with a point source for almost all globular clusters and shows no sign of extended emission (e.g., Abdo et al. 2010b). While one would expect to see point sources in $\gamma$-rays in the pulsar magnetosphere models, the IC models predict a certain degree of extension. For example, Cheng et al. (2010) suggest that the IC emission size should be >10 pc, which in the case of 47 Tucanae corresponds to ≥8′ (see also their Fig. 4), larger than the upper limit (i.e., 4.8′) reported by Abdo et al. (2010b). Therefore, unless the actual diffusion coefficients are lower than those assumed in their calculations, the model of Cheng et al. (2010) may not be able to explain the compactness of the emission size. With the increased angular resolution made available by the Pass 8 data and better photon statistics, it is expected that better constraints will be provided on the emission size of the $\gamma$-ray emission.

Clapson et al. (2011) present radio data acquired by using the Effelsberg 100-m telescope in the vicinity of Terzan 5, which may impose some constraints on the IC models. The measured radio flux at 11 cm in the circle of radius 0.15° around the core of Terzan 5 (region 1 in their Fig. 1, reproduced here as Fig. 6) is (0.14±0.21) Jy (local background emission contributes about 30–40% of this flux). Assuming a magnetic field strength of $10^6$ Gauss in the cluster, and according to Eqns. (25)–(27) of Cheng et al. (2010) that assume that the same populations of electrons lose their energy by synchrotron and IC radiation, the modeled radio fluxes at 11 cm are ~10 Jy and ~6.8 Jy, within a few arc-minutes, when the seed photons in their IC model are infrared and optical photons, respectively. It may still be possible that low-energy electrons that are responsible for the synchrotron radiation below the peak at ~44 GHz diffuse further out into region 11, thereby explaining the enhanced radio emission in region 11, for which the flux is (3.86±0.34) Jy, while high-energy electrons with a much shorter diffusion length stay close to the core, up-scatter ambient photons, and give rise to the GeV emission from Terzan 5. More detailed modeling is needed to better understand the astrophysical conditions in that cluster.

Besides originating from MSPs, $\gamma$-ray emission from particles accelerated by non-accreting WDs was also suggested (Bednarek 2012). Owing to a large number of WDs, these particles may also contribute to the observed $\gamma$-rays from globular clusters. In this scenario, particles are accelerated in the inner magnetosphere of WDs, diffuse out, and produce $\gamma$-rays via IC scattering off various soft photon fields, e.g., star light, infrared radiation, and cosmic microwave background radiation.

In summary, current $\gamma$-ray and low-frequency data are discordant with some IC models, while cumulative pulsed emission from individual MSPs with fewer modeling uncertainties remains a more favorable model for MeV–GeV emission from globular clusters. That being said, the latter model does not naturally predict any correlation of $\gamma$-ray luminosity with optical and/or infrared light at the observed cluster positions (see Section 3). Whether or not both mechanisms are at work in some $\gamma$-ray-emitting globular clusters remains an open question. Detecting more MSPs in globular clusters with future radio arrays such as SKA may help to better understand the full population of MSPs, while continuous surveys by various $\gamma$-ray detectors are equally crucial, by further increasing the sample of globular clusters that emit $\gamma$-rays (for correlational studies), and by deeper observations of individual globular clusters (to constrain the $\gamma$-ray spectrum and the emission size).

## 5. TEV OBSERVATIONS OF GLOBULAR CLUSTERS

Gamma-ray emission above 100 GeV can be produced via IC scattering off various soft photon fields by relativistic leptons accelerated in the pulsar magnetosphere (Venter et al. 2009), or leptons re-accelerated in shocks generated by pulsar winds (Bednarek & Sitarek 2007). The expected spectrum of >100 GeV emission is uncertain, as the spectrum of relativistic leptons and the nature of target photons remain uncertain in this scenario. Nevertheless, observations of nearly 20 globular clusters by using several $\gamma$-ray instruments (H.E.S.S., MAGIC, and VERITAS) above 100 GeV have been performed, but no detection has been reported yet (Kabuki et al. 2007; Aharonian et al. 2009; Anderhub et al. 2009; McCutcheon et al. 2009; Abramowski et al. 2013).

Terzan 5 is truly exceptional in this regard. The Abramowski et al. (2011) announced the 7-$\sigma$ detection of >0.4 TeV emission in the vicinity of Terzan 5 (Fig. 7). They estimated the probability of chance coincidence of this new VHE source, HESS J1747–248, with an unknown active Galactic nucleus or pulsar wind nebula to be $10^{-4}$. The reported photon flux of $1.2 \times 10^{-12}$ $cm^{-2}$ $s^{-1}$ in the 0.44–24 TeV range corresponds to 1.5% of the flux of the Crab nebula in this energy range. The spectrum can be fit by a simple power law of photon index $\Gamma_\gamma = 2.5 \pm 0.3_{stat} \pm 0.2_{sys}$. Unlike the emission in the 0.1–10 GeV range, the TeV emission region is offset from the cluster core (at the 2-$\sigma$ level) and shows sign of extended feature.

Bednarek & Sobczak (2014) discuss a scenario in which a bow shock forms ahead of where Terzan 5 is moving, and accelerated particles within the globular cluster escape preferentially in the direction opposite to that of the Terzan 5 motion, generating TeV $\gamma$-rays through IC processes. Another possible leptonic scenario is that the HESS J1747–248 originates from IC emission by electrons accelerated in colliding shocks between the collective pulsar wind from the cluster and the Galactic wind, which might also explain the fact that the TeV emission site is on the way between the cluster core and the Galactic center.

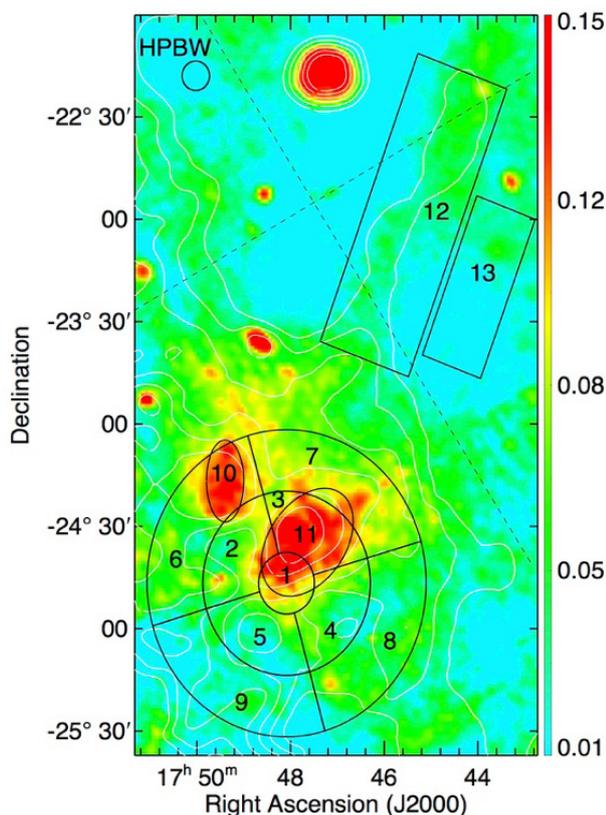

**Fig. 6.** Radio map from the Effelsberg Galactic plane surveys at 11 cm (in Jy/beam). Overlaid are white contours from the 21 cm map (for levels 0.2 to 0.7 Jy/beam). The region 1 is centered on Terzan 5 (Clapson et al. 2011).

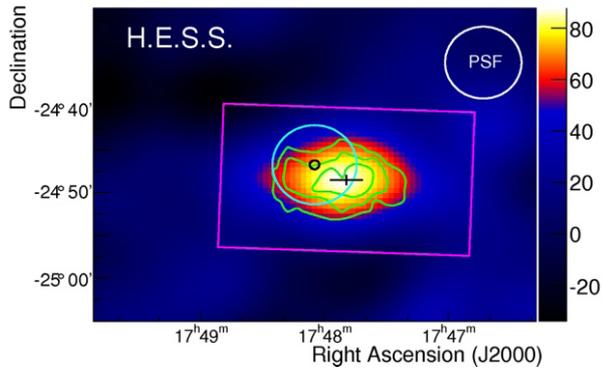

**Fig. 7.** Gamma-ray image from the HESS data. The circles show the half-mass radius (in black) and the larger tidal radius (in cyan) of the globular cluster. The cross indicates the best-fit source position of HESS J1747-248. The upper-right corner circle illustrates the instrumental PSF (Abramowski et al. 2011).

In leptonic scenarios, as noted by the Abramowski et al. (2011) and by Domainko (2011), there should be a Klein-Nishina cutoff at a few teraelectronvolts, which has not been observed. However, the lack of a cut-off may simply result from low photon statistics at these high energies, which can be tested by performing deeper TeV observations, e.g., by current IACTs or CTA.

The Abramowski et al. (2011) further discusses two hadronic scenarios: cosmic-rays accelerated by a past supernova or in a short GRB remnant. They have the advantage of being able to explain the observed simple power law spectrum up to 20 TeV. In the supernova interpretation, they argue that given the lack of molecular clouds and thus the low interstellar medium density at this location (n$\approx$0.1 $cm^{-3}$ is assumed), the energy of cosmic rays needed to produce the observed TeV flux reaches $10^{51}$ erg, which is rather high for a supernova. The more exotic short GRB scenario was further developed by Domainko (2011), but a similar energy was required and it remains unclear why this particular short GRB can be so energetic (i.e., at least $10^{51}$ erg, which is at the high end of isotropic-equivalent energy of short GRBs; Nakar 2007) and efficient in transferring the energy to cosmic-ray particles. In this scenario, thermal X-rays at the flux level of $10^{-12}$ $erg\ cm^{-2}\ s^{-1}$ are produced in shocks caused by sub-relativistic ejecta expelled during the merger event that heat the interstellar medium (Domainko & Ruffert 2008). Future X–ray observations can help to discriminate various scenarios.

In summary, the origin of HESS J1747–248 remains unclear. The simplest leptonic model for Terzan 5, namely electrons IC up-scattering off longer-wavelength emission, is strongly inconsistent with upper limits derived for 15 other globular clusters if one assumes a similar leptonic model for all these globular clusters (Abramowski et al. 2013). Future X-ray observations at the center of the TeV emission, as well as deeper TeV observations, will be crucial for tackling the origin of the TeV emission.

## ACKNOWLEDGMENTS

The authors would like to thank Meeyoung Jeong and Dmitry Olegovich Chernyshov for their help. C.Y.H. is supported by the National Research Foundation of Korea through grant 2014R1A1A2058590.

## REFERENCES

Abdo AA, Ackermann M, Ajello M, Anderson B, Atwood WB, et al., Detection of 16 Gamma-Ray Pulsars Through Blind Frequency Searches Using the Fermi LAT, Science 325, 840-844 (2009a). http://dx.doi.org/10.1126/science.1175558

Abdo AA, Ackermann M, Ajello M, Atwood WB, Axelsson M, et al., Detection of High-Energy Gamma-Ray Emission from the Globular Cluster 47 Tucanae with Fermi, Science 325, 845-848 (2009b). http://dx.doi.org/10.1126/science.1177023

Abdo AA, Ackermann M, Ajello M, Atwood WB, Axelsson M, et al., A Population of Gamma-Ray Millisecond Pulsars Seen with the Fermi Large Area Telescope, Science 325, 848-852 (2009c). http://dx.doi.org/10.1126/science.1176113

Abdo AA, Ackermann M, Ajello M, Allafort A, Antolini E, et al., Fermi Large Area Telescope First Source Catalog, Astrophys. J. Suppl. Ser. 188, 405-436 (2010a). http://dx.doi.org/10.1088/0067-0049/188/2/405

Abdo AA, Ackermann M, Ajello M, Baldini L, Ballet J, et al., A population of gamma-ray emitting globular clusters seen with the Fermi Large Area Telescope, Astron. Astrophys. 524, 75 (2010b). http://dx.doi.org/10.1051/0004-6361/201014458

Abramowski A, Acero, F, Aharonian F, Akhperjanian AG, Anton G, et al., Very-high-energy gamma-ray emission

from the direction of the Galactic globular cluster Terzan 5, Astron. Astrophys. 531, L18 (2011). http://dx.doi.org/10.1051/0004-6361/201117171

Abramowski A, Acero F, Aharonian F, Akhperjanian AG, Anton G, et al., Search for very-high-energy γ-ray emission from Galactic globular clusters with H.E.S.S., Astron. Astrophys. 551, A26 (2013). http://dx.doi.org/10.1051/0004-6361/201220719

Acero F, Ackermann M, Ajello M, Albert A, Atwood WB, et al., Fermi Large Area Telescope Third Source Catalog, Astrophys. J. Suppl. Ser. 218, 23-63 (2015). http://dx.doi.org/10.1088/0067-0049/218/2/23

Ackermann M, Ajello M, Allafort A, Atwood WB, Baldini L, et al., The First-Fermi LAT Catalog of Sources above 10 GeV, Astrophys. J. Suppl. Ser. 209, 34 (2013). http://dx.doi.org/10.1088/0067-0049/209/2/34

Aharonian F, Akhperjanian AG, Anton G, Barres de Almeida U, Bazer-Bachi AR, et al., Astron. Astrophys. 499, 273-277 (2009). http://dx.doi.org/10.1051/0004-6361/200811564

Alpar MA, Cheng AF, Ruderman MA, Shaham J, A new class of radio pulsars, Nature 300, 728-730 (1982). http://dx.doi.org/10.1038/300728a0

Anderhub H, Antonelli LA, Antoranz P, Backes M, Baixeras C, et al., Search for VHE γ-ray Emission from the Globular Cluster M13 with the Magic Telescope, Astrophys. J. 702, 266-269 (2009). http://dx.doi.org/10.1088/0004-637X/702/1/266

Bogdanov S, Grindlay JE, Heinke CO, Camilo Fernando, Freire PCC, et al., Chandra X-Ray Observations of 19 Millisecond Pulsars in the Globular Cluster 47 Tucanae, Astrophys. J. 646, 1104-1115 (2006). http://dx.doi.org/10.1086/505133

Bednarek W, Gamma-rays from millisecond pulsars in Globular Clusters, High-Energy Emission from Pulsars and their Systems, eds. Torres DF, Rea N (Springer Berlin Heidelberg, 2011), 185-205.

Bednarek W, Gamma-rays from electrons accelerated by rotating magnetized white dwarfs in globular clusters, J. Phys. G 39, 065001 (2012). http://dx.doi.org/10.1088/0954-3899/39/6/065001

Bednarek W, Sitarek J, High-energy γ-rays from globular clusters, Mon. Not. Roy. Astron. Soc. 377, 920-930 (2007). http://dx.doi.org/10.1111/j.1365-2966.2007.11664.x

Bednarek W, Sobczak T, Misaligned TeV γ-ray sources in the vicinity of globular clusters, Mon. Not. Roy. Astron. Soc. 445, 2842-2847 (2014). http://dx.doi.org/10.1093/mnras/stu1966

Chen K, Gamma-ray emission from millisecond pulsars in globular clusters, Nature 352, 695-697 (1991). http://dx.doi.org/10.1038/352695a0

Cheng KS, Taam RE, On the Origin of X-Ray Emission from Millisecond Pulsars in 47 Tucanae, Astrophys. J. 598, 1207-1216 (2003). http://dx.doi.org/10.1086/379009

Cheng KS, Chernyshov DO, Dogiel VA, Hui CY, Kong AKH, The Origin of Gamma Rays from Globular Clusters, Astrophys. J. 723, 1219-1230 (2010). http://dx.doi.org/10.1088/0004-637X/723/2/1219

Clapson AC, Domainko W, Jamrozy M, Dyrda M, Eger P, Multi-wavelength environment of the Galactic globular cluster Terzan 5, Astron. Astrophys. 532, A47 (2011). http://dx.doi.org/10.1051/0004-6361/201015559

Clark GW, X-ray binaries in globular clusters, Astrophys. J. 199, L143 (1975). http://dx.doi.org/10.1086/181869

Domainko WF, Finding short GRB remnants in globular clusters: the VHE gamma-ray source in Terzan 5, Astron. Astrophys. 533, L5 (2011). http://dx.doi.org/10.1051/0004-6361/201117538

Domainko WF, Ruffert M, Remnants of compact binary mergers, Adv. Space Res. 41, 518-522 (2008). http://dx.doi.org/10.1016/j.asr.2006.11.030

Freire PCC, Abdo AA, Ajello M, Allafort A, Ballet J, et al., Fermi Detection of a Luminous γ-Ray Pulsar in a Globular Cluster, Science 334, 1107-1110 (2011). http://dx.doi.org/10.1126/science.1207141

Gendre B, Barret D, Webb N, Discovery of a quiescent neutron star binary in the globular cluster M 13, Astron. Astrophys. 403, L11 (2003). http://dx.doi.org/10.1051/0004-6361:20030423

Harris WE, A Catalog of Parameters for Globular Clusters in the Milky Way, Astron. J. 112, 1487 (1996). http://dx.doi.org/10.1086/118116 (2010 edition, available from: http://www.physics.mcmaster.ca/~harris/mwgc.dat)

Hui CY, Cheng KS, Taam RE, Dynamical Formation of Millisecond Pulsars in Globular Clusters, Astrophys. J. 714, 1149-1154 (2010). http://dx.doi.org/10.1088/0004-637X/714/2/1149

Hui CY, Cheng KS, Wang Y, Tam PHT, Kong AKH, et al., The Fundamental Plane of Gamma-ray Globular Clusters, Astrophys. J. 726, 100-106 (2011). http://dx.doi.org/10.1088/0004-637X/726/2/100

Ivanova N, Low-Mass X-Ray Binaries and Metallicity Dependence: Story of Failures, Astrophys. J. 636, 979-984 (2006). http://dx.doi.org/10.1086/498054

Johnson TJ, Guillemot L, Kerr M, Cognard I, Ray PS, et al., Broadband Pulsations from PSR B1821-24: Implications for Emission Models and the Pulsar Population of M28, Astrophys. J. 778, 106-117 (2013). http://dx.doi.org/10.1088/0004-637X/778/2/106

Kabuki S, Enomoto R, Bicknell GV, Clay RW, Edwards PG, et al., CANGAROO III Search for Gamma Rays from Centaurus A and the ω Centauri Region, Astrophys. J. 668, 968-973

(2007). http://dx.doi.org /10.1086/520767

Katz JI, Two kinds of stellar collapse, Nature 253, 698-699 (1975). http://dx.doi.org/10.1038/253698a0

Kong AKH, Hui CY, Cheng KS, Fermi Discovery of Gamma-ray Emission from the Globular Cluster Terzan 5, Astrophys. J. Lett. 712, L36 (2010). http://dx.doi.org/10.1088/2041-8205/712/1/L36

Manandhar RP, Grindlay JE, Thompson DJ, New EGRET/COMPTEL studies of globular clusters., Astron. Astrophys. Suppl. Ser. 120, 255-260 (1996).

Manchester RN, Hobbs GB, Teoh A, Hobbs M, The Australia Telescope National Facility Pulsar Catalogue, Astron. J. 129, 1993-2006 (2005). http://dx.doi.org/10.1086/428488

McCutcheon M, for the VERITAS Collaboration, VERITAS Observations of Globular Clusters, in 31st International Cosmic Ray Conference (ICRC), Lodz, Poland, 7-15 Jul 2009.

McLaughlin DE, Anderson J, Meylan G, Gebhardt K, Pryor C, et al., Hubble Space Telescope Proper Motions and Stellar Dynamics in the Core of the Globular Cluster 47 Tucanae, Astrophys. J. Supple. Ser. 166, 249-297 (2006). http://dx.doi.org/10.1086/505692

Michelson PF, Bertsch DL, Brazier K, Chiang J, Dingus BL, et al., EGRET upper limits to the high-energy gamma-ray emission from the millisecond pulsars in nearby globular clusters, Astrophys. J. 435, 218-224 (1994). http://dx.doi.org/10.1086/174808

Nakar E, Short-hard gamma-ray bursts, Phys. Rep. 442, 166-236 (2007). http://dx.doi.org/10.1016/j.physrep.2007.02.005

Nolan PL, Abdo AA, Ackermann M, Ajello M, Allafort A, et al., Fermi Large Area Telescope Second Source Catalog, Astrophys. J. Suppl. Ser. 199, 31-76 (2012). http://dx.doi.org/10.1088/0067-0049/199/2/31

Ortolani S, Barbuy B, Bica E, Zoccali M, Renzini A, Distances of the bulge globular clusters Terzan 5, Liller 1, UKS 1, and Terzan 4 based on HST NICMOS photometry, Astron. Astrophys. 470, 1043-1049 (2007). http://dx.doi.org/10.1051/0004-6361: 20066628

Pooley D, Lewin WHG, Anderson SF, Baumgardt H, Filippenko AV, et al., Dynamical Formation of Close Binary Systems in Globular Clusters, Astrophys. J. Lett. 591, L131 (2003). http://dx. doi.org/10.1086/377074

Stappers BW, Archibald AM, Hessels JWT, Bassa CG, Bogdanov S, et al., A State Change in the Missing Link Binary Pulsar System PSR J1023+0038, Astrophys. J. 790, 39-47 (2014). http://dx.doi.org/10.1088/0004-637X/790/1/39

Strong AW, Moskalenko IV, Propagation of Cosmic-Ray Nucleons in the Galaxy, Astrophys. J. 509, 212-228 (1998). http://dx.doi.org/10.1086/306470

Takata J, Li KL, Leung GCK, Kong AKH, Tam PHT, et al., Multi-wavelength Emissions from the Millisecond Pulsar Binary PSR J1023+0038 during an Accretion Active State, Astrophys J. 785, 131-137 (2014). http://dx.doi.org/10.1088/0004-637X/785/2/131

Tam PHT, Kong AKH, Hui CY, Cheng KS, Li C, et al., Gamma-ray Emission from the Globular Clusters Liller 1, M80, NGC 6139, NGC 6541, NGC 6624, and NGC 6752, Astrophys. J. 729, 90-97 (2011). http://dx.doi.org/10.1088/0004-637X/729/2/90

Tam PHT, Li KL, Kong AKH, Takata J, Leung GCK, et al., The birth of radio millisecond pulsars and their high-energy signature, in the 10th INTEGRAL Workshop, Annapolis, MD, 15-19 Sep 2014.

Venter C, de Jager OC, Constraining A General-Relativistic Frame-Dragging Model for Pulsed Radiation from a Population of Millisecond Pulsars in 47 Tucanae using GLAST LAT, Astrophys. J. Lett. 680, L125 (2008). http://dx.doi.org/10.1086/589996

Venter C, de Jager OC, Clapson AC, Predictions of Gamma-Ray Emission from Globular Cluster Millisecond Pulsars Above 100 MeV, Astrophys. J. Lett. 696, L52 (2009). http://dx.doi.org/10.1088/0004-637X/696/1/L52

Verbunt F, Hut P, The Globular Cluster Population of X-Ray Binaries, The Origin and Evolution of Neutron Stars, IAU Symposia 125, eds. Helfand DJ, Huang JH (Reidel Publishing Co., Dordrecht, 1987), 187-197.

Wu JHK, Hui CY, Wu EMH, Kong AKH, Huang RHH, et al., Search for Pulsed $\gamma$-Ray Emission from Globular Cluster M28, Astrophys. J. Lett. 765, L47 (2013). http://dx.doi.org/10.1088/2041-8205/765/2/L47